\documentclass[conference]{IEEEtran}
\makeatletter
\def\endthebibliography{%
  \def\@noitemerr{\@latex@warning{Empty `thebibliography' environment}}%
  \endlist
}
\makeatother

\usepackage{array}
\newlength\mylength
\newcolumntype{P}[1]{>{\centering\arraybackslash}p{#1}}

\ifCLASSINFOpdf
  \usepackage[pdftex]{graphicx}
\else
   \usepackage[dvips]{graphicx}
\fi
\usepackage{amsmath}
\usepackage{bm}
\usepackage{soul,color}
\usepackage{lipsum}
\usepackage{comment}
\usepackage{algorithm}  
\usepackage{algpseudocode}  

\makeatletter
\newcommand\fs@betterruled{%
  \def\@fs@cfont{\bfseries}\let\@fs@capt\floatc@ruled
  \def\@fs@pre{\vspace*{2pt}\hrule height.8pt depth0pt \kern2pt}%
  \def\@fs@post{\kern2pt\hrule\relax}%
  \def\@fs@mid{\kern2pt\hrule\kern2pt}%
  \let\@fs@iftopcapt\iftrue}
\floatstyle{betterruled}
\restylefloat{algorithm}
\makeatother

\usepackage[backend=biber,
sorting=none,
doi=false,
isbn=false,
url=false,
maxnames=3, minnames=2,
style=ieee, ]{biblatex}
\begin{document}
\bstctlcite{IEEEexample:BSTcontrol}
%
\title{LiDAR Aided Human Blockage Prediction for 6G}
%
%
%
\author{
\IEEEauthorblockN{ Dileepa Marasinghe, Nandana Rajatheva and Matti Latva-aho }\\
\IEEEauthorblockA{\textit{Centre for Wireless Communications,} \\
\textit{Univeristy of Oulu,}\\
Oulu, Finland \\
E-mail: dileepa.marasinghe@oulu.fi,
nandana.rajatheva@oulu.fi, matti.latva-aho@oulu.fi}

}

\pagenumbering{gobble}

\maketitle

\begin{abstract}
Leveraging higher frequencies up to THz band paves the way towards a faster network in the next generation of wireless communications. However, such shorter wavelengths are susceptible to higher scattering and path loss forcing the link to depend predominantly on the line-of-sight (LOS) path. Dynamic movement of humans has been identified as a major source of blockages to such LOS links. In this work, we aim to overcome this challenge by predicting human blockages to the LOS link enabling the transmitter to anticipate the blockage and act intelligently. We propose an end-to-end system of infrastructure-mounted LiDAR sensors to capture the dynamics of the communication environment visually, process the data with deep learning and ray casting techniques to predict future blockages. Experiments indicate that the system achieves an accuracy of 87\% predicting the upcoming blockages while maintaining a precision of 78\% and a recall of 79\% for a window of 300 ms.
\end{abstract}


%
\IEEEpeerreviewmaketitle

\section{Introduction}

Communication networks in the 6G era are envisioned to utilize the enormous bandwidth available in the higher frequencies ranging from mmWave to THz. However, communicating in these bands is extremely challenging as they suffer from high scattering due to their smaller wavelengths compared to conventionally used sub-six GHz frequencies. To combat this, ultra-narrow beams are used and as the beam-widths get smaller the random blockages to the LOS path can degrade the link quality abruptly  \cite{tan2020thz}. Specifically,  blockage occurring with dynamic movements of the humans in the environment has been reported as a significant degrading factor of the link quality \cite{human_blockage}. One solution to minimize this adverse effect of link quality degradation is to predict such blockages allowing the transmitter to act proactively by doing a handover to a transmitter with LOS availability or avoid transmitting in the blocking time duration. Predicting such events require a holistic view of the communication environment. Sensing the environment through heterogeneous modalities emerge as a promising technology to capture the dynamics of the communication environment implying information on the scattering, thus aiding in the communication procedures such as positioning, beam management and blockage prediction. Cameras, LiDARs and radars are potential sensors that can be used for such environment sensing. Particularly, LiDAR sensors are an interesting choice to capture the 3D structure of the communication environment in a detail-rich manner which also eliminates the privacy concerns in contrast to cameras. The use of LiDARs for such monitoring purposes was restrained by the enormous price of the sensors. However, recent developments in the LiDAR technology have reduced the price to couple of hundred dollars which enables a multitude of applications to harness the potential of LiDARs. Moreover, the range of these sensors become adequate to capture the communication environment since the 6G networks converge to tiny cells in terms of coverage distance because of the high path loss in the higher frequencies. Furthermore, algorithms for processing such hetero-modal data to generate predictive insights, specially for vision data has evolved significantly in the recent past with the continuous growth of deep learning techniques.

Therefore, in this work we try to look at the problem of dynamic blockage prediction from the emerging field of vision aided communication point of view. We propose a system to utilize infrastructure-mounted LiDAR sensors \cite{NalinEliD, padmal2021elevated} to predict human blockage in indoor scenarios. We solely rely on the LiDAR data eliminating the burden to the wireless algorithms and provide positioning data and future blockage status of the users in the communication environment.

The rest of the paper is organized as follows. In section II, we summarize the recent efforts to utilize vision data for communications and work on blockage prediction. We formulate the problem in section III and describe our proposed system for the human blockage prediction problem in section IV. Section V contains the details on the experimental setup we used. We present the results and a discussion on the simulations conducted in section VI and conclude the paper in section VII with our intentions of future work.

\section{Previous work}

Use of visual data from sensors such as LiDAR and camera  for aiding wireless communications is relatively a new area that has gained attention recently. In \cite{Aldebaro}, a deep learning based mmWave beam-selection problem in a vehicular  scenario has been explored for a downlink system with analog beamforming. The authors focus on solving LOS versus non-LOS (NLOS) classification of the current channel and selection of top-M beam pairs for the beamforming from a predefined codebook and shows the potential of using LiDAR data for aiding mmWave communication procedures. The authors in \cite{Dias} consider a similar problem to \cite{Aldebaro} using LiDAR and position data for a mmWave downlink vehicular scenario and conclude that a distributed architecture  where LiDAR processing is done at vehicles performs better.

\begin{figure*}
    \centering
    \includegraphics[width=0.9\textwidth]{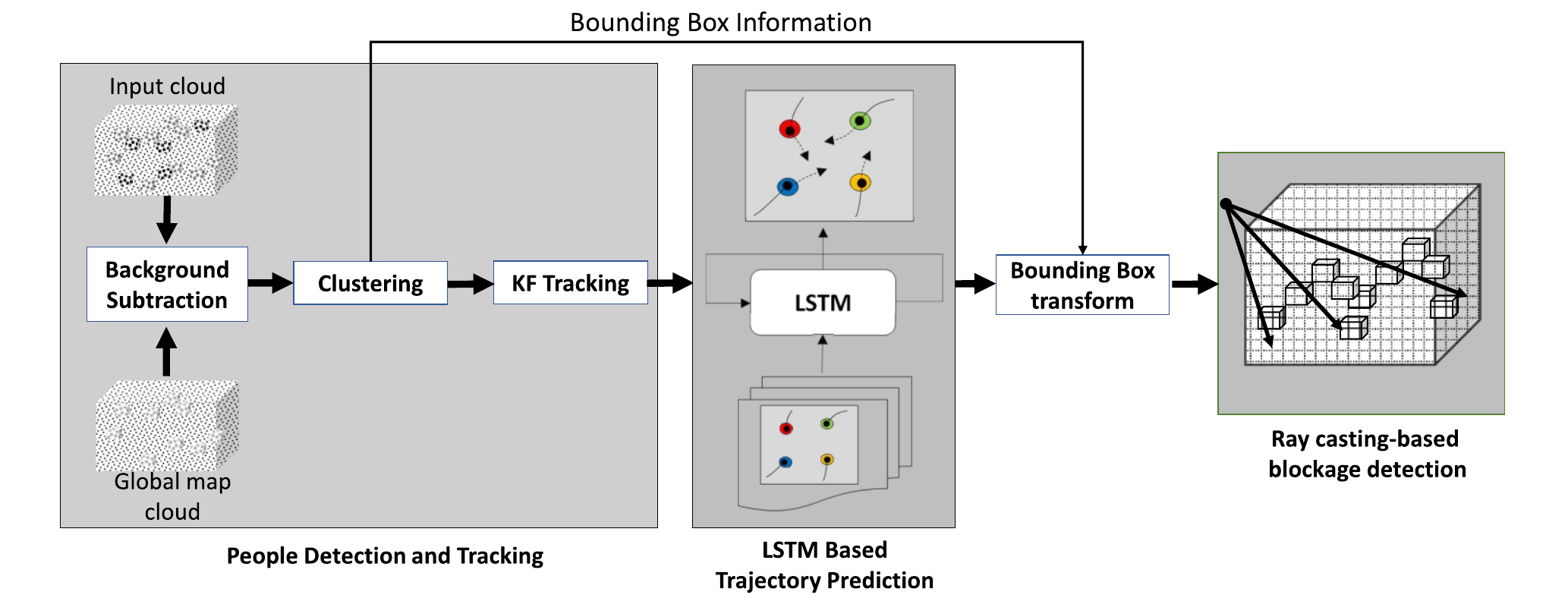}
    \caption{System overview}
    \label{fig:system}
\end{figure*}

Apart from using LiDAR data, use of images from fixed cameras has been explored for mmWave beam selection in \cite{Xu} and \cite{charan2020visionaided}. In \cite{Xu} a panoramic point cloud generation using camera images is carried out in contrast to obtaining point cloud data from LiDAR. The results conclude that the panoramic point clouds constructed by the proposed method are more appropriate for accurate beam selection when compared with the local point cloud scanned by user as in previous works. This emphasizes the fact that having a wider view of the environment improves the beam selection accuracy. 

Base station selection based on human blockage prediction for a static transmitter receiver pair has been studied utilizing RGBD cameras in \cite{proactivebasestation}. The dynamic blockage prediction problem for mmWave networks in a vehicular scenario has been explored in \cite{charan2020visionaided} and extended to proactive handover problem in \cite{charan2021visionaided}. The authors use camera images from a camera fixed on the base station to predict whether the user will be blocked. A recurrent neural network with gated recurrent units (GRU) is trained with camera images and beam indices. For object detection a pre-trained deep learning network is used. The proposed approach outperforms the method using beam sequences only demonstrating successful use of visual information to aid in wireless algorithms. 

\section{Problem formulation }

The main objective of this work is to utilize the LiDAR data to predict the dynamic blockage in a future window. We entirely depend on the data from the LiDARs to determine the blockage events, thus radio blockage is directly evaluated as the visual blockage. We consider an indoor area where humans are walking randomly and all the humans are utilizing a user equipment (UE) to connect to an access point (AP) inside the same area operating in mmWave/THz frequencies. The radio links between the transmitter and the UEs are assumed to be established already. Multiple LiDAR sensors are mounted in elevated positions to monitor the indoor area which capture the 3D information and sent to a server which processes the data to generate a combined 3D map of the indoor area at a rate of $f_s$ synchronized with the sensor rates. We denote the link status of the radio link between the AP and the UE carried by a human $p$ at a time instance $t$ as $l_p^{t}$. If there exist a LOS path for the link, $l_p^{t} = 0$ and if blocked which means in NLOS condition, $l_p^{t} = 1$. Note that the time is considered in discrete form synchronized with the publishing of the 3D maps. We define the link status of the UE of human $p$, for a prediction window of $w$ time instances as,
\begin{gather*}
l_p^{t_w} =
\begin{cases}
   0 &  l_p^{t} = 0 \quad  \forall t \in [{t_{e} + 1}, t_{p}]   \\    
   1 & otherwise. 
\end{cases}
\end{gather*}
The objective of the method is to  process the combined 3D maps in an observation window of $t \in [{t_{s}}, t_{e}]$ time instances, and evaluate the link statuses $l_p^{t_w}$ of the UEs in the prediction window $t \in [{t_{e} + 1}, t_{p}]$.  

\section{LiDAR aided human blockage prediction}

We propose an end-to-end system for the blockage prediction problem utilizing point cloud processing and deep learning. The solution consists of three stages. Fig. 1 shows an overview of the system. First stage processes the received 3D maps for the detection of humans in the environment and tracking their movements. The second stage predicts the trajectories of the tracked humans based on the past accumulated position data. The third stage synthesizes the future instance with the bounding boxes and predicted positions of the humans and detect potential blockages using a minimal raytracing algorithm.  Next we describe the three stages in the following subsections.

\subsection{Point cloud based human detection and tracking}
Human detection and tracking stage is based on the proposed method in \cite{koide} for static LiDAR sensors. First, offline environmental mapping is done using the infrastructure-mounted LiDAR system for identifying the static environment and stored as a point cloud which we term as the global map. During the operation, point clouds from the sensors are registered to generate the 3D map using the method described in \cite{padmal2021elevated}. Then the background points are removed by subtracting the global map from the current 3D map to identify the dynamic points. The dynamic points are subjected to Euclidean clustering to detect potential human clusters. This clustering process might result in inaccuracies if the humans are close to the each other which is mitigated using the Haselich’s split-merge clustering algorithm \cite{Haselich}. The detected human clusters are assumed to be walking on the ground plane, thus the tracking is done using $x,y$ coordinates discarding the $z$ coordinate. A Kalman filter with a constant velocity model  and global nearest neighbor data association are used for tracking the detected human clusters between the frames.

Let $N$ be the number of people tracked at the time instance $t=t$. The tracking algorithm outputs the  position and bounding boxes of a tracked human $p$ as a tuple $\bm{D}^t_p=[ \bm{x}^t_p, \bm{b}^t_p] $ where $\bm{x}^t_p = (x_p,y_p,z_p)$ is the position in Cartesian coordinates and $\bm{b}^t_p$ is the bounding box. We consider two methods with axis aligned bounding boxes (AABB) and oriented bounding boxes (OBB). An AABB is defined as $\bm{b}^t_p = ( x_{min},y_{min}, z_{min}, x_{max},y_{max},z_{max})$ and an OBB as $\bm{b}^t_p = ( \hat{x}_{min},\hat{y}_{min},\hat{z}_{min}, \hat{x}_{max},\hat{y}_{max},\hat{z}_{max}, \bm{M})$ where $\hat{x}$ denotes coordinates in object space and $\bm{M}$ is the rotation matrix with respect to the global coordinate system. The calculation of OBB for a human cluster is carried out by computing eigen vectors of the points in the cluster to determine the orientation and calculation of AABB is straightforward. 

\subsection{Trajectory prediction}
Human motions are complex which adds an extra difficulty in predicting their trajectories compared to vehicles, robots etc. Predicting a motion path of a human can be viewed as a sequence generation  problem when the past trajectory of the human is known. Recent works report considerable success in applying recurrent neural networks (RNN), particularly long short term memory (LSTM) models for human trajectory forecasting \cite{sociallstm,trajnet}. LSTM models have shown the ability to encompass the sequential nature while adapting to the non-linearities in human motion. Following the state of the art, we use an LSTM model for predicting the human trajectory based on the tracking data provided by the previous stage. Each human trajectory is modelled by one LSTM which will encompass the motion of the human in the observation time window and the future trajectory is predicted as a sequence using the LSTM. As the humans are assumed to be walking in the ground plane, similar to first stage, we consider only the $x,y$ coordinates of $\bm{x}^t_p$ for a trajectory as $\bm{\Tilde{x}}^t_p = ( x_p , y_p )$. Therefore, the observed trajectory for a human $p$ is $\bm{R}_p = \{\bm{x}^t_p\}^{t_{e}}_{t = t_{s}} $ in the observation window $t= [{t_{s}}, t_{e}]$. We  estimate the future trajectory of the human $p$, $\bm{\Hat{R}}_p = \{\bm{\Tilde{x}}^t_p\}^{t_{p}}_{t = t_{e} + 1}$ for the prediction window  $t= [{t_{e} + 1}, t_{p}]$ using the LSTM model as follows. 

Each position of a trajectory in the observation window are converted to velocities in each dimension by subtracting the previous position from the current position and sent as input to the LSTM model which gives the following recurrence:

\begin{subequations}
\begin{align}
e^t_p &= \phi ( \Delta \bm{\Tilde{x}}^t_p; W_{emb}), \\
h^t_p &= LSTM ( h^{t-1}_p, e^t_p ; W_{enc}),
\end{align}
\end{subequations}where $\phi$ is and embedding function with ReLU non-linearity and $W_{emb}, W_{enc} $ are the embedding and encoder weights of the model. Then the velocity distribution of the time step, $t+1$ is estimated using the hidden state of the time step, $t$ for each of the time steps in the prediction window. Similar to \cite{sociallstm},\cite{trajnet} a bi-variate Gaussian distribution is considered as the velocity distribution with mean $\mu_p^{t+1} = ( \mu_x , \mu_y )$, variance $\sigma_p^{t+1} = (\sigma_x , \sigma_y)$ and correlation coefficient $\rho_p^{t+1}$. This is learnt through:

\begin{equation}
    [\mu_p^{t+1}, \sigma_p^{t+1}, \rho_p^{t+1}] = \phi_{dec}( h^{t}_p; W_{dec}),
\end{equation}
with $\phi_{dec}$ is modelled with a fully connected layer having $W_{dec}$ as the weight matrix. Training of the LSTM model is done with human trajectories using negative log-likelihood loss. The future trajectory of the human $\bm{\Hat{R}}_p$ is generated by sampling the resulting bi-variate Gaussian for each of the time steps in the prediction window as:

\begin{equation}
    \bm{\Tilde{x}}^{t+1}_p =  \bm{\Tilde{x}}^{t}_p + \bm{v},  
\end{equation}
where $\bm{v} \sim \mathcal{N} ( \mu_p^{t+1} , (\sigma_p^{t+1}, \rho_p^{t+1}))$.

\subsection{Ray casting based blockage detection}

\begin{figure}[ht]
    \centering
    \includegraphics[width=\linewidth]{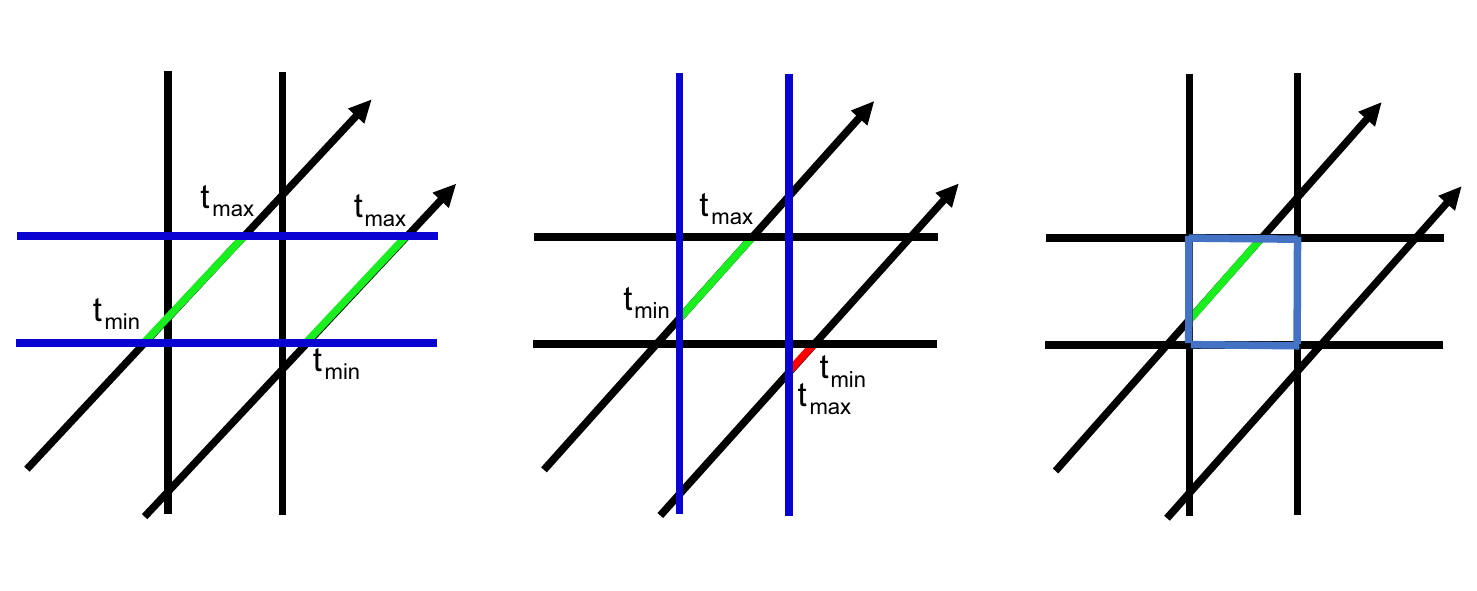}
    \caption{Slab method for ray-box intersection in 2D. Left figure shows the two rays clipped by horizontal planes with intersection points $t_{min}$ and $t_{max}$. The middle figure shows the next step of clipping by vertical planes and taking the maximum between current $t_{min}$ and new intersection points and minimum between current $t_{max}$ and new intersection points. Finally a check is done to determine the intersection whether $t_{max} > t_{min}$. Right figure shows that only the intersecting ray has a remaining portion inside the box.}
    \label{fig:slabmethod}
\end{figure}

This stage utilizes the predicted trajectories and current tracking information and bounding boxes to evaluate the blockage in the predicted positions using ray casting. Ray casting is a technique used in computer graphics rendering which utilizes geometric relations to evaluate the interactions of a ray originating from an emitter with the objects in the considered scene. In this work we utilize the ray-box intersection algorithm for the AABBs known as the slab method \cite{raybox}. The key idea in the slab method is to consider an AABB as three pairs of parallel planes. The intersection of the ray with the bounding box is evaluated by clipping the ray with each pair of the parallel planes and determine whether any portion of the ray is remaining which means there is an intersection of the ray with the AABB which is elaborated in Fig. \ref{fig:slabmethod}. For OBBs the slab method is applied in object coordinate space instead of the global coordinate space  as the bounding box is an AABB in the object space although the rotation is with respect to global space. The considered ray is transformed from the global space to object space using the position and the rotation matrix of the OBB. 

The predicted trajectories of the humans, $\bm{\Hat{R}_p}$ from the second stage and the associated tracking information  $\bm{D^t_p}$ from the first stage are taken as input to this stage.  For the blockage evaluation, we transform the bounding boxes to the predicted positions in each time step in the prediction window. Then the ray originating from the transmitter point to the center point of a transformed bounding box is considered. We evaluate whether each of these rays intersect with the bounding boxes of the humans tracked  using the method outlined in Algorithm 1 based on the slab method. If there are intersections detected for a particular ray, we check whether the closest intersection to the transmitter is for the considered bounding box or any other. If the closest intersection is not with the bounding box of the human in consideration, then we declare that the human is blocked which means in NLOS condition.

\begin{algorithm}[!t]
1: \textbf{Input} :  \{($\bm{D^}{t_e}_p$, $\bm{\Hat{R}}_p)\}_{p=1}^N$ \\
2: \textbf{Output} : $\{l_p^{t_w}\}_{p=1}^N$ \\
3: \textbf{Initialize} : $ l_p^{t_w} \leftarrow 0\  \forall p$  \\
4: \textbf{for} $t \leftarrow t_{e} + 1 , t_p$ \textbf{do} \\
5: \quad $d_{min}^p \leftarrow Inf \ \forall p$ \\
6: \quad $i_{min}^p \leftarrow p \ \forall p$ \\
7: \quad \textbf{for} $p \leftarrow  1, N$ \textbf{do} \\
8: \quad \quad $\bm{\overline{b}}^t_p \leftarrow BoundingBoxTransform (\bm{b}^{t_e}_p,\bm{\Tilde{x}}^t_p  ) $ \\
9: \quad \quad  \textbf{for} $q= 1 $ to $N$ \textbf{do} \\
10: \quad \quad \quad $\textbf{if}: l_q^{t_w} = 0$ and $\bm{\Tilde{x}}_q^t \boldsymbol{\cdot} \bm{\Tilde{x}}_p^t > 0 $\\
11: \quad \quad \quad \quad \quad \quad \quad \quad \quad and $\| \bm{\Tilde{x}}_q^t - \bm{\Tilde{x}}_{tx}\| < \| \bm{\Tilde{x}}_p^t - \bm{\Tilde{x}}_{tx}\| $\\
12: \quad \quad \quad \quad \quad $d_{intersect} \leftarrow CheckIntersect(\bm{x}_{tx}, \bm{\Tilde{x}}_q^t, \bm{\overline{b}}^t_p)$\\
13: \quad \quad \quad \quad\quad  \textbf{if}: $d_{intersect} < d_{min}^q$ \textbf{then}\\
14: \quad \quad \quad \quad \quad  \quad $d_{min}^q \leftarrow d_{intersect}$\\
15: \quad \quad \quad \quad \quad  \quad $i_{min}^q \leftarrow p$\\
16: \quad \quad \quad \quad\quad  \textbf{end if} \\
17: \quad \quad \quad $\textbf{end if}$  \\
18: \quad \quad \textbf{end for}\\
19: \quad \textbf{end for}\\
20: \quad\textbf{for} $p \leftarrow  1, N$ \textbf{do} \\
21: \quad \quad \textbf{if}: $i_{min}^p \neq p$ \textbf{then} \\
22: \quad \quad  \quad $l_p^{t_w} \leftarrow 1$\\
23: \quad \quad  \textbf{end if} \\
24: \quad \textbf{end for}\\
25: \textbf{end for}\\
\caption{Ray casting based blockage detection}
\label{Algorithm1}
\end{algorithm}

In Algorithm 1, the function $BoundingBoxTransform$ computes the translation and rotation of a human from current time step to a predicted time step based on the predicted positions and computes the bounding box at the predicted time step. The $CheckIntersect$ function performs ray-box intersection test and returns the distance form the transmitter to the intersection point using the slab method when AABBs are considered. In case of OBBs, $CheckIntersect$ first transforms the coordinates of the transmitter and destination point of the ray to its object space based on the transformed bounding box and perform  ray-box intersection test in object coordinate space. Since the identification of blockages in all the time steps within the window is an exhaustive search, we apply few enhancements to reduce the number of searches per each human in the prediction window in Algorithm 1. Although the outermost loop starting in line 4 evaluates the scenario for each time step, the first condition at line 10 checks whether a previous time step has detected blockage. If a blockage has been detected, then the algorithm skips checking any other time steps for the ray directed towards that human, as we are interested in determining any blockage during a prediction window. Furthermore the other two conditions in lines 10 and 11 ensure that the ray-box intersection check is done only for humans who are closer to the transmitter than the considered human and both have similar directional vectors with respect to the transmitter. 

\section{Experimental setup}


\subsection{Data generation method}
\begin{figure}[ht]
    \centering
    \includegraphics[width=0.9\linewidth]{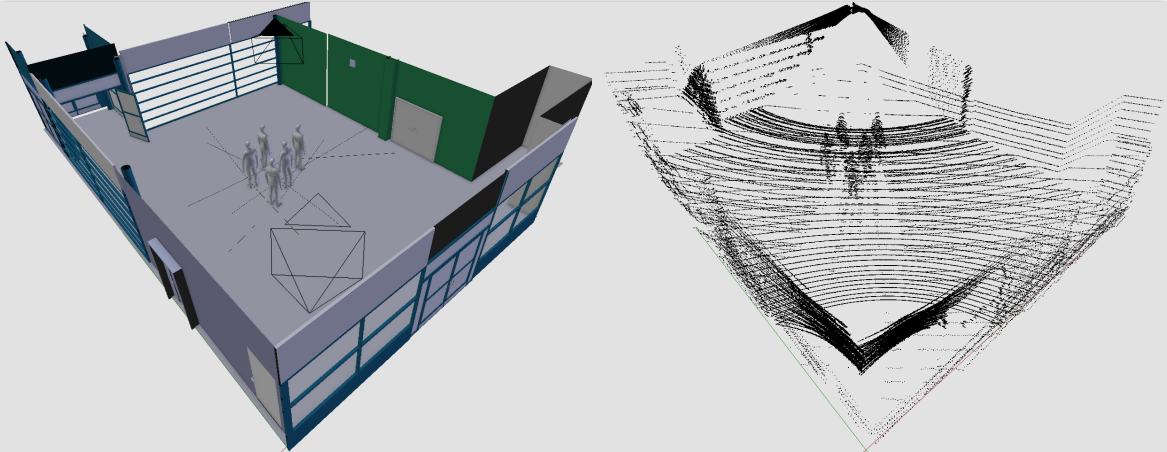}
    \caption{An example instance in Blensor LiDAR data generation. Left figure is the 3D modelled environment in Blender and the right figure shows generated LiDAR point clouds based on the 3D model.}
    \label{fig:blensor}
\end{figure}

We use simulations for evaluating the proposed system in this work. LiDAR point clouds are simulated using Blensor \cite{blensor} which is an open source software tool for simulating LiDAR and depth sensors with 3D animations based on Blender \cite{blender} which is a computer graphics animation tool. Blensor provides the facility of adding LiDAR/depth sensors on a Blender based animation and generating the point clouds as the 3D animation progresses. We generate 2D human trajectories using ORCA simulator \cite{orca} as utilized in \cite{trajnet}. Trajectories are generated such that the humans start randomly from a point in the circumference of a 12.5 m radius circle and walk towards the antipodal position on the circle. We generate 500 scenes with 10 humans starting at random positions and a human model is animated to follow these trajectories in the 3D environment in Blensor while point clouds are generated from the deployed sensors. We used two Velodyne HDL-32E LiDAR sensors with 50 m range operating at 10 Hz. Fig. \ref{fig:blensor} shows an example 3D scene modeled in Blensor where the sensors are placed on the elevated positions, tilted 35 degrees downwards to cover the indoor area and the resulting LiDAR point clouds. We extract the 3D transformation matrices to register the point clouds from the two sensors and generate the input point cloud to the system. For the LSTM model, we use the generated trajectory data for training and validation with a 60-40 split using the implementation provided for \cite{trajnet}. Blockage status ground truth data for each human in a scene are generated for each frame using the raytracing functionality in the 3D model in Blender. 

\subsection{Implementation}

The implementation of the system was carried out on the Robot Operating System (ROS) which is a software tool platform for building robots. 
The three stages  outlined in the system as in Fig. \ref{fig:system} are launched separately while the data transfer is done thought ROS messages. The human  detection and tracking stage implementation is an adapted version of the open source code available for \cite{koide}. We generate the globalmap cloud from Blensor without any humans and load it to the system prior to the start. The input point clouds are published to the system by emulating the sensors from the stored point clouds in the data generation stage. 
The detection and tracking nodes publish the positions and bounding boxes of the humans in the scene which is subscribed by the trajectory predictor node running the second stage. For the LSTM model, we used an embedding layer dimension of 64 and LSTM  hidden state dimension of 512 which was trained using the 2D data generated from ORCA simulator for 30 epochs. 
The inference is done utilizing a 10-frame sliding window for past trajectory data provided from the detection and tracking data. The predicted trajectories are sent to the ray casting node which evaluates blockage status of each tracked human in the prediction window.


\section{Results and Discussion}
\begin{figure}[ht]
    \centering
    \includegraphics[width=0.6\linewidth]{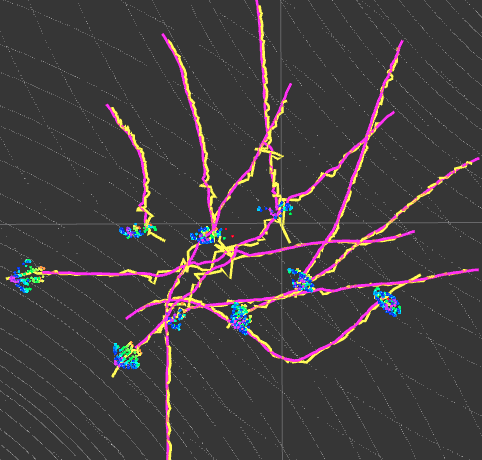}
    \caption{Top view visualization of the trajectory prediction. Magenta lines show the ground truth path while yellow lines are the predicted paths based on LSTM method.}
    \label{fig:trajectory}
\end{figure}

In this section we present the results of the proposed method based on the simulation data generated. We evaluated the proposed system with 500 scenes from the test dataset generated. Visual inspection of the  trajectory prediction results as in Fig. \ref{fig:trajectory} shows satisfactory agreement with the ground truth trajectories. It can be seen that the LSTM model successfully adapts to the non-linear trajectories of the humans. Average displacement error (ADE) in the test dataset with the future time step is given in Table \ref{table1}  for the trained LSTM model. The error increases as the prediction time step increases.

\begin{table}
\caption{Average displacement error with future time step (metres) }\label{table1}
\begin{tabular}{|P{\mylength}|P{\mylength}|P{\mylength}|P{\mylength}|P{\mylength}|P{\mylength}|P{\mylength}|P{\mylength}|P{\mylength}|}
 \hline
  1 & 2 & 3 & 4 & 5 & 6 & 7 & 8 & 9 \\
  \hline
 0.22 & 0.27 & 0.34 & 0.42 & 0.51 & 0.61 & 0.72 & 0.83 & 0.96 \\ 
 \hline
\end{tabular}
\end{table}

\begin{figure}[t]
    \centering
    \includegraphics[width=\linewidth]{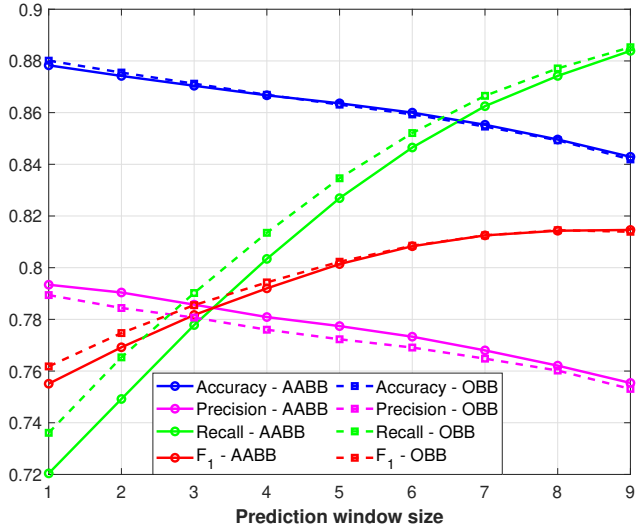}
    \caption{Evaluation metric scores vs. window size}
    \label{results}
\end{figure}

The blockage prediction problem is a binary classification problem in the future window. Therefore, we evaluated our proposed methods with evaluation metrics for binary classification. The accuracy is defined as the total number of correct results from all the samples given which means correctly identified LOS windows and NLOS windows out of all the instances considered. As shown in Fig. \ref{results}, both AABB and OBB methods achieve an accuracy around 88\% and for shorter prediction windows and decrease monotonically as we try to predict further into the future since the ADE increases with the time steps. This problem is inherently an unbalanced problem since there will be more LOS instances than NLOS instances which means accuracy itself is not sufficient to evaluate the method. Therefore, we look at the metrics precision, recall and $F_1$ score for more insights. $F_1$ score gives the harmonic mean between precision and recall and a high $F_1$ score means a better predictor.
For this blockage prediction scenario, precision indicates the fraction of blockage windows correctly predicted from the all the predicted blockage windows while recall indicates the fraction of blockage windows correctly predicted from the true blocked windows. 



Clearly, the recall in both methods increases as the window size increases in Fig. \ref{results}. The reason is because the prediction window is considered blocked if there is at least one blockage event during the window. As the window size increases, the probability of having blockage instances in the window increases which results in fewer false negatives. Concurrently, the precision drops as the number of false positives increase for the same reason. The $F_1$ score settles down around 0.815 when the window size is increased to 8 for both methods.
When comparing the two methods, OBB method outperforms the AABB method in recall although precision is higher in AABB method and both methods have very similar accuracy values. However, $F_1$ score is higher with OBB method which means it performs better. Nevertheless, OBB method involves higher complexity when transforming all the rays in to the object space due to the matrix inversions used in the transformations. Therefore, with a trade-off between complexity and performance, AABB method can be considered as the better option. The preferable window size for the prediction can be chosen as 3 which corresponds to 300 ms when the system is running at 10 Hz as both methods have the intersection of precision and recall around window size  of 3. 

We observed tracking failures are a significant source of error in the proposed system. We used a Kalman filter and point cloud processing to detect and track the humans which can further be improved with deep learning-based detection and tracking of humans. Moreover, we used the Vanilla LSTM model to predict a human trajectory. As humans move with respect to their dynamics of the surroundings incorporating interactions between the humans can improve the performance of the trajectory prediction \cite{trajnet} which should improve the blockage prediction accuracy. Furthermore, we completely relied on the visual data in this work, and incorporating information from the wireless algorithms such as received power has the ability to increase the prediction accuracy further which will be considered in our future work. 

\section{Conclusions}
In this work, we proposed a system for dynamic blockage prediction in an indoor scenario utilizing visual information from infrastructure-mounted LiDAR sensors. Our method consisted of 3 stages with point cloud processing, trajectory prediction, and raycasting. We demonstrated that the LiDAR data can be used to predict incoming blockages with a satisfactory performance which will aid the higher frequency systems which will be used in 6G. The proposed system achieved an accuracy 87\% with a precision of 78\% and a recall of 79\% for a prediction window of 300 ms. In future work, evaluating the proposed system with real LiDAR data and the use of data from the wireless system to improve the solution will be explored. 


%

\ifCLASSOPTIONcaptionsoff
  \newpage
\fi



%
\printbibliography

%

\end{document}